\input{aipcheck}

\documentclass[
    draft             % use draft while you are working on the paper
  ]
  {aipproc}

\layoutstyle{6x9}

\newcommand{\ET}      {\ensuremath{E_{\mathrm{T}}}}
\newcommand{\Etiso}   {\ensuremath{E_{\mathrm{T}}^{\mathrm{iso}}}}

\newcommand{\squishlist}{
 \begin{list}{\textbullet}
  { \setlength{\itemsep}{0pt}
     \setlength{\parsep}{3pt}
     \setlength{\topsep}{3pt}
     \setlength{\partopsep}{0pt}
     \setlength{\labelwidth}{0.5em}
     \setlength{\labelsep}{0.5em} } }

\newcommand{\squishlisttwo}{
 \begin{list}{$\bullet$}
  { \setlength{\itemsep}{0pt}
     \setlength{\parsep}{0pt}
    \setlength{\topsep}{0pt}
    \setlength{\partopsep}{0pt}
    \setlength{\leftmargin}{2em}
    \setlength{\labelwidth}{1.5em}
    \setlength{\labelsep}{0.5em} } }

\newcommand{\squishend}{
  \end{list}  }

\begin{document}

\title{Measurements of isolated prompt photons in $pp$ collisions at 7 TeV in ATLAS}


\author{Giovanni Marchiori (LPNHE Paris)\\ for the ATLAS Collaboration}{
  address={giovanni.marchiori@lpnhe.in2p3.fr}
}

%\author{for the ATLAS Collaboration}

\begin{abstract}
Two recent measurements of the cross section for the inclusive
production of isolated prompt photons in $pp$ collisions at a
center-of-mass energy $\sqrt{s} = 7$ TeV are presented. 
The results are based on data collected in 2010 with the ATLAS detector
at the Large Hadron Collider. 
The measurements cover the pseudorapidity ranges $|\eta^\gamma|<1.37$
 and $1.52\leq |\eta^\gamma|<2.37$ and the
transverse energy range $15 \leq E_T^\gamma < 400$ GeV.
The measured cross sections are compared to predictions from
next-to-leading-order perturbative QCD calculations. 
\end{abstract}

\maketitle

%%%%%%%%%%%%%%%%%%%%%%%%%%%%%%%%%%%%%%%%%%%%
%% MAINMATTER
%%%%%%%%%%%%%%%%%%%%%%%%%%%%%%%%%%%%%%%%%%%%

\section{Introduction}

Prompt photon production at hadron colliders proceeds mainly through
parton hard scattering, thus providing a handle for testing
perturbative QCD (pQCD) predictions. 
% In particular, the measurement of the inclusive production cross section
% does not require to reconstruct additional jets in the event and the
% differential cross section as a function of the photon energy can
% therefore provide precise constraints.
%~\cite{Angelis_promptphoton,QCD_promptphoton_NLO}.    
The dominant production mechanism at Large Hadron
Collider (LHC) energies is $qg{\rightarrow}q\gamma$,
thus its cross section measurement can constrain
the gluon density in protons.
%~\cite{Akesson_promptphoton,promptphoton_and_gluon_pdf}.

Here we present two measurements
%~\cite{promptphotonpaper_2010,promptphoton_confnote_2011} 
of the inclusive isolated prompt photon production cross section, as a
function of the photon transverse energy, using $pp$ collision data
collected in 2010 with the ATLAS detector at the LHC at a
center-of-mass energy of 7 TeV. The
former~\cite{promptphotonpaper_2010} is based on an integrated 
luminosity $\int \mathcal{L} dt = (0.88\pm 0.1)$~pb$^{-1}$, and
provides a measurement of the cross section for $15\leq E_T^\gamma
<100$~GeV in the photon pseudorapidity intervals $[0,0.6)$,
$[0.6,1.37)$ and $[1.52,1.81)$. The
latter~\cite{promptphoton_confnote_2011} uses the 
full 2010 data sample ($\int\mathcal{L}dt = (34.6\pm 1.2)$ pb$^{-1}$), 
covers the transverse energy range $45\leq E_T^\gamma<400$~GeV and
explores an additional pseudorapidity interval, $[1.81,2.37)$.
% An isolation requirement is applied to both reduce the dominant
% background from photon pairs produced in decays of neutral mesons in
% jets.
% and to reduce the contribution to the total cross section from
% partons fragmentation into photons, which is the part that is
% theoretically less known.

\section{The ATLAS detector}
The ATLAS detector~\cite{ATLAS_detector,ATLAS_CSC} consists of an
inner tracking system surrounded by a thin superconducting solenoid
providing a 2 T magnetic field, electromagnetic and hadronic
calorimeters, and a muon spectrometer based on large superconducting
toroids. 
For the measurements presented here, the calorimeter and the inner
detector are of particular relevance.
The former provides a precise measurement of the
photon energy and direction, and discrimination between single photons
and fake photon candidates.
% (typically merged photon pairs from decays
%of neutral mesons inside jets). 
The latter is used to distinguish electrons from photons and to
reconstruct conversion vertices. 
These two systems and the photon trigger algorithms are briefly
described in Ref.~\cite{promptphotonpaper_2010}.

\section{Event selection}
%
% trigger
%
Events are triggered using a single-photon high-level trigger with
a nominal transverse energy threshold of 10
GeV~\cite{promptphotonpaper_2010} or 40
GeV~\cite{promptphoton_confnote_2011}.
Using unbiased or lower-threshold triggers, these triggers
are found to be fully efficient for prompt photons passing the
selection criteria described later.
%
% data quality
%
Events in which the calorimeters or the inner detector are not fully 
operational, or show data quality problems, are discarded.
%
% non-collision bkg
%
To reduce non-collision backgrounds, events are required to
have at least one reconstructed primary vertex, with at 
least three associated tracks, consistent with the average beam spot.

%
% photon selection
%
Photons are reconstructed from electromagnetic clusters and tracking
information provided by the inner detector as described
in~\cite{promptphotonpaper_2010}.
Photon candidates near regions of the calorimeter affected by read-out or
high-voltage failures are not considered. 
Events with one photon candidate in the nominal acceptance are selected:
\squishlist
%\begin{itemize}
\item $\ET^\gamma>15$~GeV, $|\eta^\gamma|< 1.37$ or $1.52\leq
|\eta^\gamma|<1.81$ in~\cite{promptphotonpaper_2010},
\item $\ET^\gamma>45$~GeV, $|\eta^\gamma|< 1.37$ or $1.52\leq
|\eta^\gamma|<2.37$ in~\cite{promptphoton_confnote_2011}.
%\end{itemize}
\squishend
%In case of multiple photon candidates satisfying these requirements we
%retain the {\em leading}-\ET~one ({\em i.e.} the photon with highest
%transverse energy).
Background from non-prompt photons originating from decays
of energetic $\pi^{0}$ and $\eta$ mesons inside jets are 
suppressed by means of shower-shape and
isolation variables.
Photon candidates are required to pass {\em tight} identification 
criteria based on nine discriminating variables computed from the
the lateral and longitudinal profiles of the energy deposited
in the calorimeters, in particular in the first (finely
segmented) and second layers of the electromagnetic calorimeter
and in the hadronic calorimeter behind it.
The photon {\em transverse isolation energy} \Etiso\ is computed from the
sum of the energies in the electromagnetic calorimeter cells in a cone
of radius 0.4 in the $\eta-\phi$ plane around the photon axis. The
contribution to \Etiso\ from the photon itself and from the soft-jet
activity from the underlying event are subtracted (for the latter the
technique proposed in~\cite{Cacciari:UE} is adopted).
We require $\Etiso<3$~GeV.
The final sample size is 110 thousand events in~\cite{promptphotonpaper_2010}
and 174 thousand events in~\cite{promptphoton_confnote_2011}. 
About 30\% of the photon candidates are reconstructed from conversions.

\section{Background subtraction}
A non-negligible residual contribution of background candidates is
expected in the selected photon sample.
The main source is due to misidentified QCD jets, typically
containing a neutral meson ($\pi^{0}$, $\eta$) that carries
most of the jet energy and decays to a collimated photon pair.
The background contamination is estimated and then subtracted by means
of a data-driven counting technique based on the observed number of
events in the control regions ({\em sidebands}) of a two-dimensional
plane formed by the photon transverse isolation energy and a photon
identification variable, exploiting two properties: the negligible correlation
between these two variables for background events and the dominance of
background over signal in the three background control regions. 
The number of signal events $N_A^{\rm sig}$ in the selected sample is given by
$N_A^{\rm sig} = N_A - (N_B-c_BN_A^{\rm sig})\frac{(N_C-c_CN_A^{\rm
sig})}{(N_D-c_DN_A^{\rm sig})}$,
where $N_A$ is the total number of events in the selected sample,
$N_K$ (for $K \in \{B,C,D\}$) are the number of events in the three
control regions and $c_K\equiv\ N^{\rm sig}_K/ N^{\rm sig}_A$
are {\em signal leakage fractions}, extracted
from simulated signal events (their size does not exceed a few \%).
The procedure is applied separately for each of the pseudorapidity
intervals under study and in several bins of photon transverse
energy. The (small) background contribution from isolated electrons
from $W$ and $Z$ is estimated from simulated $W/Z$ events using the
$e\to\gamma$ fake rate $(\approx 8\%)$ measured in data.
The estimated signal purity $N_A^{\rm sig}/N_A$ increases
from around 50\% at $\ET=15$~GeV to 90\% and above for $\ET>100$~GeV.
It is cross-checked with a template fit to the isolation distribution
of the selected photons.

\section{Cross section measurement}
The \ET-differential cross section for each of the pseudorapidity
intervals under study is determined from the estimated signal
yield as a function of the photon transverse energy and the photon
trigger ($\varepsilon_{\mathrm{trig}}$), reconstruction
($\varepsilon_{\mathrm{rec}}$) and identification
($\varepsilon_{\mathrm{ID}}$) efficiencies, 
 $\frac{d\sigma}{dE_{\rm T}^{\gamma}} = \frac{N_{A}^{\rm sig}~U}{\left(\int{\mathcal{L}dt}\right)~\Delta{E_{\rm T}^{\gamma}}~\varepsilon_{\mathrm{trig}}~\varepsilon_{\mathrm{rec}}~\varepsilon_{\mathrm{ID}}}.$
While the trigger efficiency (consistent with 100\%) is measured in data, the
other two are estimated from simulated signal samples, after adjusting
the shower shape distributions in order to match those in data. A less precise
data-driven measurement of $\varepsilon_{\mathrm{ID}}$ using electrons
from $W$ decays is in agreement with the nominal
result. 
$\varepsilon_{\mathrm{rec}}$ is around 80\%\footnote{A significant
inefficiency was due to dead readout regions and was
recovered in winter 2011 shutdown}, while
$\varepsilon_{\mathrm{ID}}$ increases from around 65\% to $\approx
95\%$ as a function of $\ET^\gamma$.
Correction factors $U$ (close to 1) take into account
migrations between neighbouring bins in $\ET$ due to energy
resolution. They are obtained with various {\em unfolding} techniques~\cite{bayes_unfolding,SVD_unfolding},
using the $\ET^{\rm true}\leftrightarrow\ET^{\rm rec}$ response matrix
from simulated true photons.

%\section{Systematic uncertainties}
Several sources of systematic uncertainties are evaluated and their
contributions are combined, taking into account their correlations.
The uncertainty on the reconstruction efficiency is dominated by the
isolation efficiency cut (3-4\%), the signal event generator (2\%) and
the limited knowledge of the detector material (1-2\%). The uncertainty
on the identification efficiency is dominated by the detector
material knowledge (up to 6\%) and the data/simulation shower shape
agreement (up to 5\%). The uncertainty on the background subtraction is
dominated by the inputs from the simulation (up to 10\%) and the
choice of the background control regions (up to 6\%). The
uncertainty on the photon energy scale (1.5-3\%) translates into a
global 5-10\% uncertainty on the cross section.

\section{Results and conclusion}
\begin{figure}[!htbp]
  \includegraphics[height=.28\textheight]{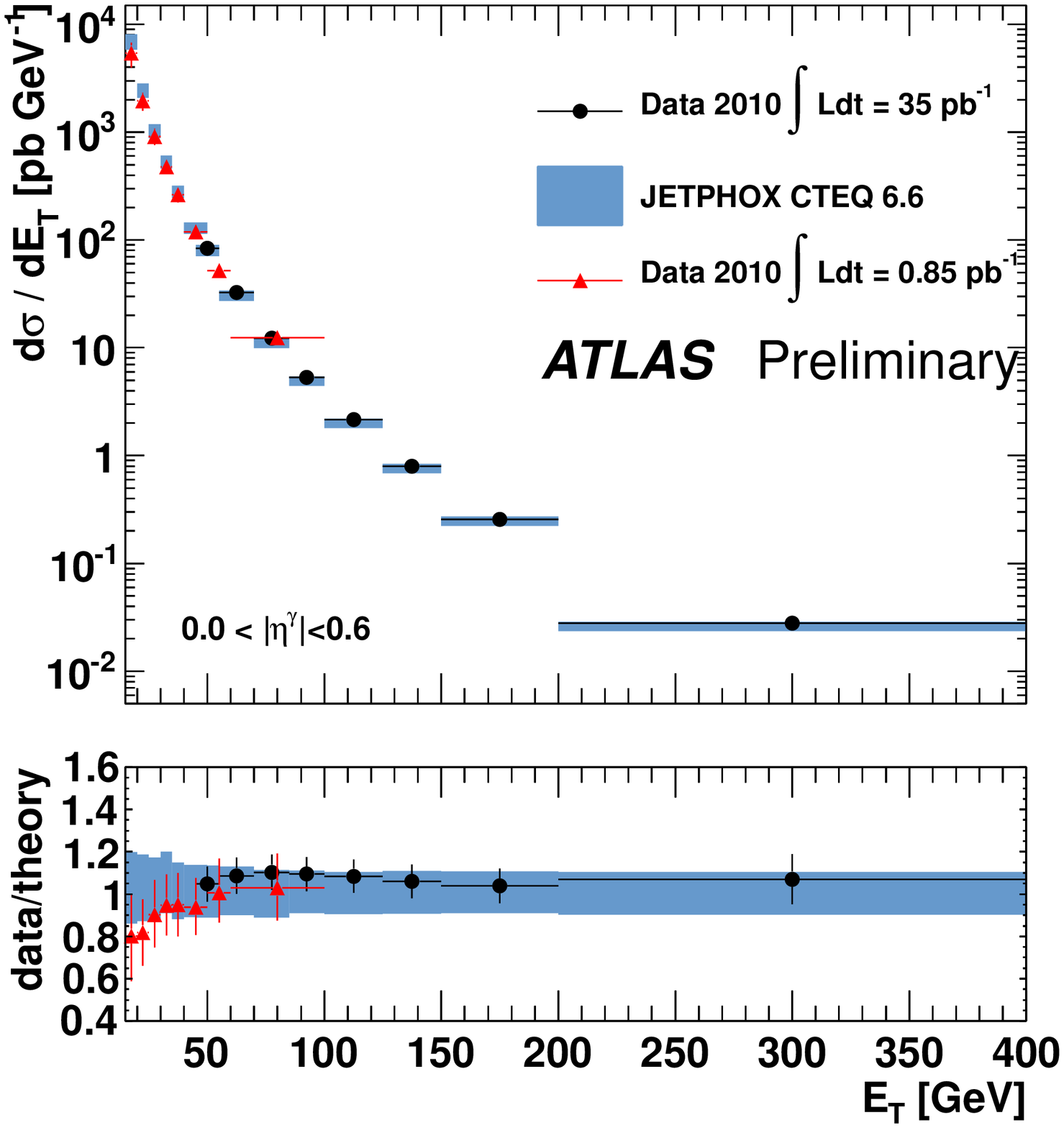}
  \includegraphics[height=.28\textheight]{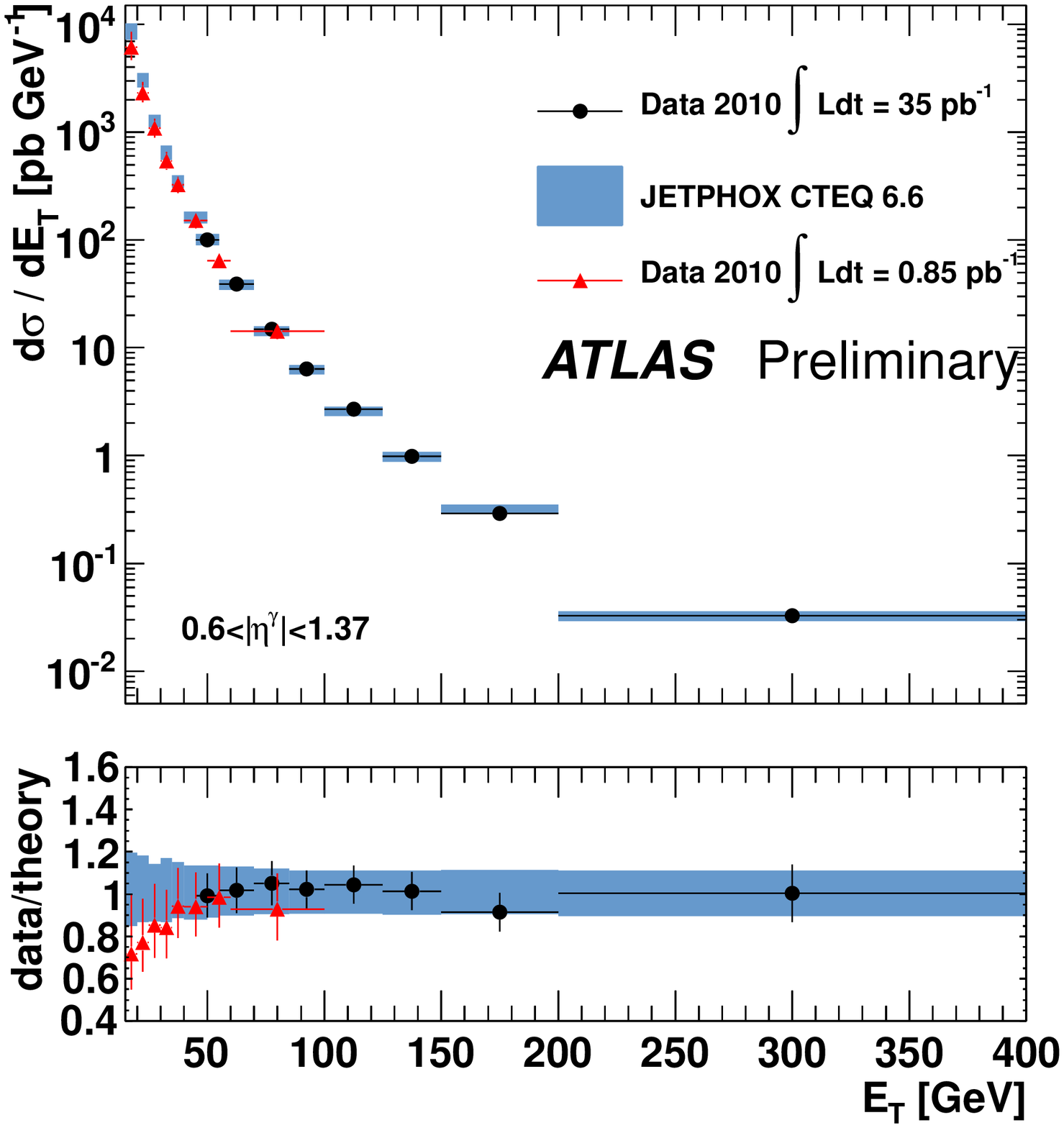}
  \caption{Prompt $\gamma$ production cross sections
  for $|\eta^\gamma|{<}0.6$ (left) and $0.6{\leq}|\eta^\gamma|{<}1.37$ (right).} 
  \label{fig:xsec_barrel}
\end{figure}

\begin{figure}[!htbp]
  \includegraphics[height=.28\textheight]{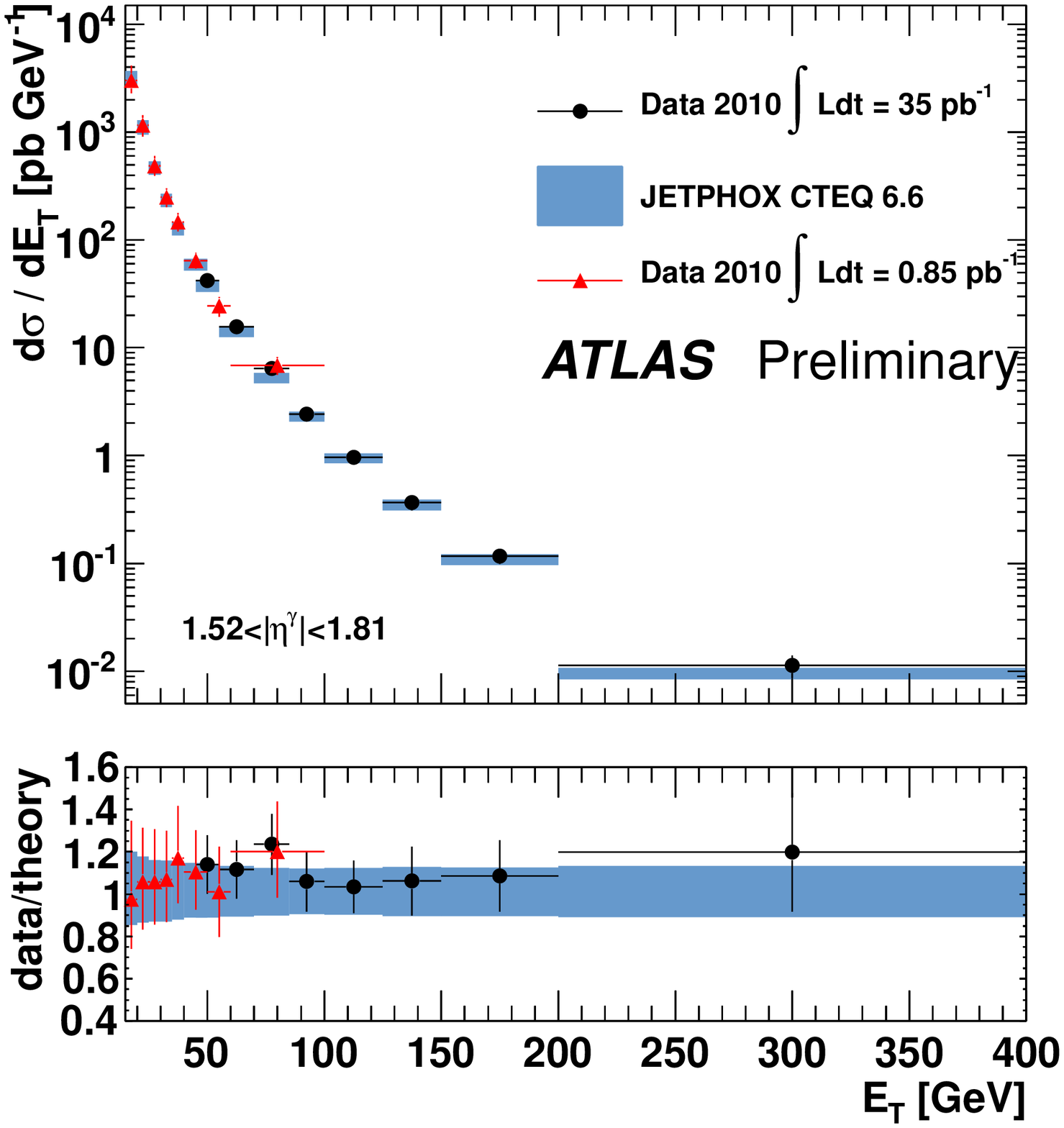}
  \includegraphics[height=.28\textheight]{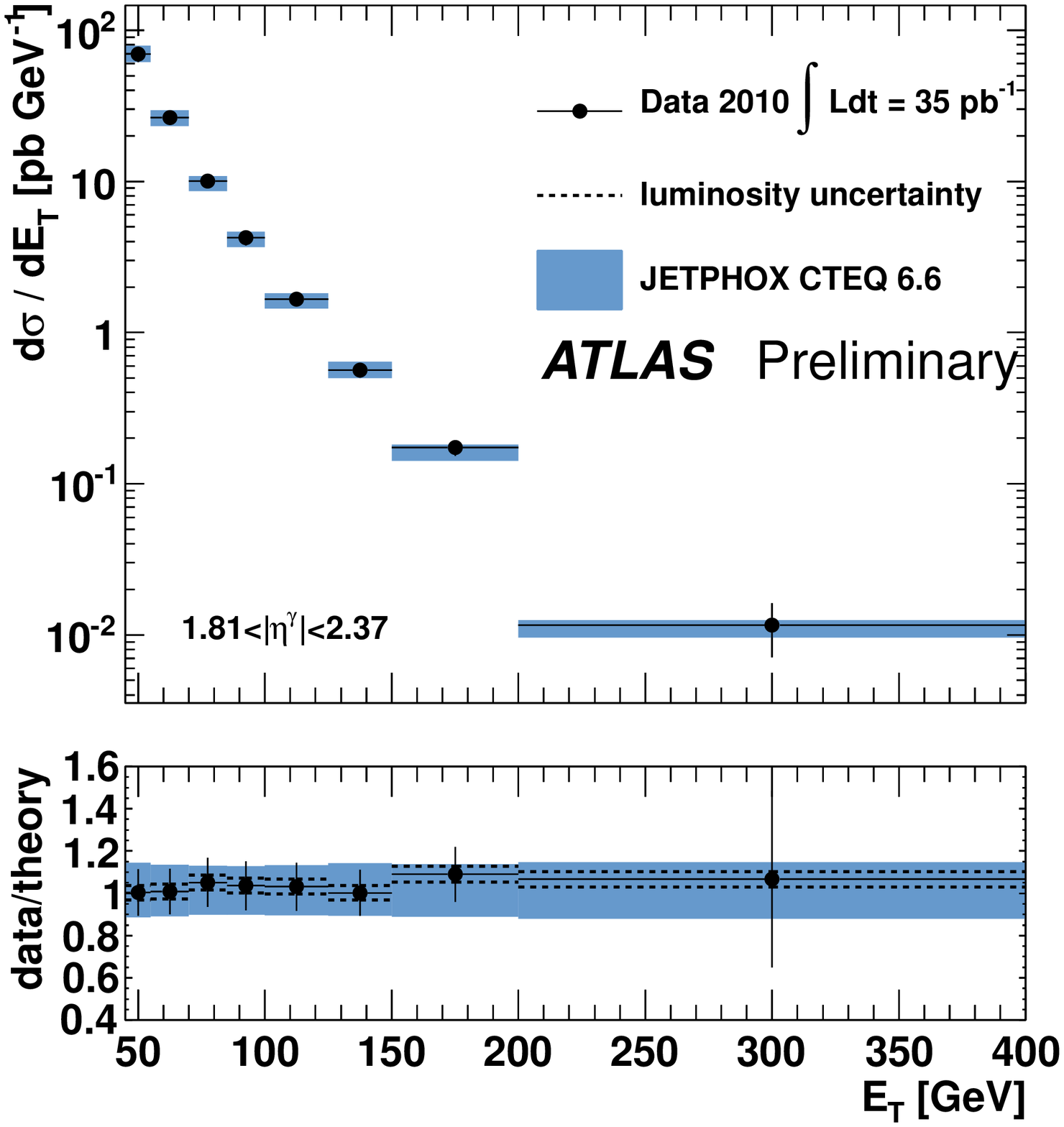}
  \caption{Prompt $\gamma$ production cross sections
  for $1.52{\leq}|\eta^\gamma|{<}1.81$ (left) and $1.81{\leq}|\eta^\gamma|{<}2.37$ (right).} 
  \label{fig:xsec_endcap}
\end{figure}

The measured cross sections are shown in Fig.~\ref{fig:xsec_barrel}
and~\ref{fig:xsec_endcap}. The red triangles represent the
experimental results from~\cite{promptphotonpaper_2010}, the black
dots those from~\cite{promptphoton_confnote_2011}. The theoretical
pQCD cross section, computed with a fixed-order NLO parton-level
generator~\cite{jetphox}, is overlaid (blue band).
The CTEQ 6.6 PDFs are used. The
scales are set to $\ET^\gamma$. The parton transverse
energy in a cone of radius 0.4 around the photon is required to be
below 4 GeV: varying the requirement by $\pm 2$ GeV changes
the cross section by $\pm 2\%$.
PDF uncertainties lead to a systematic uncertainty decreasing from 4\% to
2\% with \ET, and varying the scales
between 0.5 \ET~and 2 \ET~leads to an uncertainty decreasing
from 20\% to 8\%. 


%\section{Conclusion}
The measured cross sections are in agreement between the two
measurements~\cite{promptphotonpaper_2010,promptphoton_confnote_2011}
and with the theoretical predictions~\cite{jetphox} for
$\ET^\gamma>35$ GeV. Below, where the contribution from parton-to-photon
fragmentation is larger, the theory tends to overestimate
the data, possibly hinting to the need of more accurate (NNLO) predictions.

%%%%%%%%%%%%%%%%%%%%%%%%%%%%%%%%%%%%%%%%%%%%%%%%
%% BACKMATTER
%%%%%%%%%%%%%%%%%%%%%%%%%%%%%%%%%%%%%%%%%%%%%%%%

%\begin{theacknowledgments}
%\end{theacknowledgments}



\bibliographystyle{aipproc}   % if natbib is available

\bibliography{marchiori}

\hyphenation{Post-Script Sprin-ger}
\begin{thebibliography}{8}
\expandafter\ifx\csname natexlab\endcsname\relax\def\natexlab#1{#1}\fi
\providecommand{\enquote}[1]{``#1''}
\expandafter\ifx\csname url\endcsname\relax
  \def\url#1{\texttt{#1}}\fi
\expandafter\ifx\csname urlprefix\endcsname\relax\def\urlprefix{URL }\fi
\providecommand{\eprint}[2][]{\url{#2}}

\bibitem[Aad et~al.(2011)]{promptphotonpaper_2010}
G.~Aad, et~al., \emph{Phys.~Rev.~D} \textbf{83}, 052005 (2011).

\bibitem[{ATLAS Collaboration}(2011)]{promptphoton_confnote_2011}
{ATLAS Collaboration}, \emph{ATLAS-CONF-2011-058}  (2011).

\bibitem[Aad et~al.(2008)]{ATLAS_detector}
G.~Aad, et~al., \emph{JINST} \textbf{3}, S08003 (2008).

\bibitem[Aad et~al.(2009)]{ATLAS_CSC}
G.~Aad, et~al., \emph{arXiv:0901.0512 [hep-ex]}  (2009).

\bibitem[Cacciari et~al.(2010)]{Cacciari:UE}
M.~Cacciari, G.~P. Salam, and S.~Sapeta, \emph{JHEP} \textbf{04}, 065 (2010).

\bibitem[D'Agostini(1995)]{bayes_unfolding}
G.~D'Agostini, \emph{Nucl.~Instrum.~Methods~A} \textbf{362}, 487 (1995).

\bibitem[Hoecker and Kartvelishvili(1996)]{SVD_unfolding}
A.~Hoecker, and V.~Kartvelishvili, \emph{Nucl.~Instrum.~Methods~A}
  \textbf{372}, 469 (1996).

\bibitem[Fontannaz et~al.(2001)]{jetphox}
M.~Fontannaz, J.~P. Guillet, and G.~Heinrich, \emph{Eur. Phys. J.}
  \textbf{C21}, 303--312 (2001).

\end{thebibliography}

\IfFileExists{\jobname.bbl}{}
 {\typeout{}
  \typeout{******************************************}
  \typeout{** Please run "bibtex \jobname" to optain}
  \typeout{** the bibliography and then re-run LaTeX}
  \typeout{** twice to fix the references!}
  \typeout{******************************************}
  \typeout{}
 }

\end{document}